\documentclass[pra,aps,epsfig,psfig,multicols,showpacs,tightenlines]{revtex4}

\usepackage{graphics,bm}
\usepackage{graphicx}

\newcommand{\beq}{\begin{equation}}
\newcommand{\eeq}{\end{equation}}
\newcommand{\bqa}{\begin{eqnarray}}
\newcommand{\eqa}{\end{eqnarray}}
\def\gsim{\mathrel {\vcenter {\baselineskip 0pt \kern 0pt
\hbox{$>$} \kern 0pt \hbox{$\sim$} }}}

\protect

\begin{document}
\title{Exact Solitonic Solutions of the One-Dimensional Gross-Pitaevskii Equation with a
Time-Dependent Harmonic Potential and Interatomic Interaction}
\author{Usama Al Khawaja}
\affiliation{ \it Physics Department, United Arab Emirates
University, P.O. Box 17551, Al-Ain, United Arab Emirates.}

\date{\today}

\begin{abstract}

We derive exact solitonic solutions of the one-dimensional time-dependent Gross-Pitaevskii equation with
time-dependent strengths of the harmonic external potential and the interatomic interaction. The time-dependence
of the external potential and interatomic interaction are given in terms of a general function of time. For an
oscillating strength of the external potential, the solutions correspond to breathing single and multiple
solitons. The amplitude and frequency of the oscillating potential can be used to control the dynamics of the
center of mass of the solitons. For certain values of these parameters, the solitons can be {\it trapped} at the
center of the harmonic potential.

\end{abstract}

\pacs{}

\maketitle

\section{ Introduction}

Since the experimental realization of solitons in Bose-Einstein condensates
\cite{burger,denschlag,anderson,randy,schreck,eirman}, intense interest in their properties has emerged
\cite{linear1,linear2,cast,sala,busch,abdu,usamapre}. Recently, exact solitonic solutions of the time-dependent
Gross-Pitaevskii equation that describes the behavior of the condensate, have been obtained
\cite{wu,jun,liang,lu,raj,usamapre}. We have shown in \cite{usama_darboux}, that exact solutions of the
Gross-Pitaevskii equation can not be obtained for general external potential and interatomic interaction. For
the Gross-pitaevskii equation to be solved exactly, the external potential and the interatomic interaction must
be related. As an example on such a relation, the external potential in the Gross-Pitaevskii equation solved in
Ref.~\cite{liang} is an expulsive harmonic potential and the interatomic interaction has an exponential
prefactor that grows with time with a rate that is related to the strength of the harmonic potential. Our
investigation in \cite{usama_darboux} has provided a mathematical proof for the existence of such a correlation
between the external potential and the interatomic interaction.

From an experimental point of view, it would be more interesting to find exact solutions for the external
potentials and interatomic interactions used in the experiments. For instance, it would be more interesting to
have exact solutions for the realistic case of an upright harmonic external potential and constant interatomic
interaction rather than for an inverted harmonic potential and exponentially growing interatomic interaction
strength as in \cite{liang}. It turns out, however, that for the Gross-Pitaevskii equation to be solved exactly
with an upright harmonic potential, the interatomic interaction will have to be oscillatory and divergent at
regular times \cite{usama_darboux}.

In this paper, we use the Darboux transformation method \cite{salle} to investigate the possibility of obtaining
exact solutions of a time-dependent Gross-Pitaevskii equation with strengths of the harmonic external potential
and interatomic interaction that are general functions of time. It turns out that some special cases of such a
general form of harmonic potential and interatomic interaction can be easily realized experimentally by
controlling the strength of the magnetic field that provides the harmonic trap and using Feshbach management of
interatomic interaction \cite{fesh}. One interesting special case corresponds to an oscillating strength of the
harmonic potential and practically constant strength of the interatomic interaction. The amplitude and frequency
of the oscillating external potential can be used to delay the {\it escape} of the soliton from the harmonic
potential or even to {\it trap} it at the center.

The rest of the paper is organized as follows. In the next section, we present the general form of the
Gross-Pitaevskii equation to be solved. In section~\ref{darboux} we use the Darboux transformation method to
derive the exact solutions. In section~\ref{results}, we present and discuss the properties of the solutions in
the special case of an oscillating external potential. We end in section~\ref{conc} with a summary of our main
conclusions.

\section{The Gross-Pitaevskii equation}
\label{gpsec}

When the confinement of the Bose-Einstein condensate is much larger
in, say the $y$ and $z$ (transverse) directions, compared to the
confinement in the $x$ direction, the system can be considered
effectively one-dimensional along the $x$ direction. The
three-dimensional Gross-Pitaevskii equation can then be {\it
integrated} over the $y$ and $z$ directions to result in the
following one-dimensional Gross-Pitaevskii equation
\cite{carr,usama_darboux}
\begin{equation}
i{\partial\psi(x,t)\over\partial t}=-{\partial^2\psi(x,t)\over\partial x^2}-{1\over4}\lambda
p(t)x^2\psi(x,t)-2aq(t)|\psi(x,t)|^2\psi(x,t) \label{gp1}.
\end{equation}
Here $a$ is the $s$-wave scattering length, and $\lambda=2\omega_x/\omega_\perp$, where $\omega_x$ and
$\omega_\perp$ are the characteristic frequencies of the harmonic trapping potential in the $x$ and transverse
directions, respectively. In the last equation, length is scaled to $a_{\perp}$, time to $2/\omega_\perp$, and
$\psi(x,t)$ to $1/\sqrt{2a_\perp}$, where $a_{\perp}=\sqrt{\hbar/m\omega_\perp}$ is the characteristic length of
the harmonic trap in the transverse direction. The dimensionless general functions $p(t)$ and $q(t)$ are
introduced to account for the time-dependencies of the strengths of the trapping potential and the interatomic
interaction.

For Eq.~\ref{gp1} to be solved exactly using the Darboux transformation method, the functions $p(t)$ and $q(t)$
must be parametriclly related to each other throw a general function $g(t)$ \cite{usama_darboux}. In this case,
the Gross-Pitaevskii equation, that we derive exact solutions for, takes the form
\begin{equation}
i{\partial\psi(x,t)\over\partial t}=-{\partial^2\psi(x,t)\over\partial x^2}-{1\over4}\lambda \left[\lambda
g(t)^2-{\dot g}(t)\right]x^2\psi(x,t)-2a e^{2c_1+\lambda\int{g(t)dt}}|\psi(x,t)|^2\psi(x,t) \label{gp2},
\end{equation}
where $c_1$ is an arbitrary constant.

\section{Darboux Transformation and the Exact Solutions}
\label{darboux} In the Darboux transformation method, we start by finding a linear system of equations for an
{\it auxiliary} field ${\bf \Psi}(x,t)$ such that Eq.~(\ref{gp2}) is its {\it consistency condition}
\cite{salle, usama_darboux}. We find that Eq.~(\ref{gp2}) corresponds to the consistency condition of the
following linear system \cite{note}:
\begin{equation}
{\bf \Psi}_x=\zeta{\bf J}{\bf\Psi}{\bf\Lambda}+{\bf P}{\bf \Psi}
\label{psi_x},
\end{equation}
\begin{equation}
{\bf \Psi}_t=2i\zeta^2{\bf J}{\bf \Psi}{\bf
\Lambda}^2+2i\zeta {\bf P}{\bf \Psi}{\bf \Lambda}+\lambda x
g \zeta {\bf J}{\bf \Psi}{\bf \Lambda}+{\bf W}{\bf \Psi}
\label{psi_t},
\end{equation}
where,\\
\begin{math}
{\bf \Psi}(x,t)=\left(\begin{array}{cc}
\psi_1(x,t)&\psi_2(x,t)\\
\phi_1(x,t)&\phi_2(x,t)
\end{array}\right)
\end{math},
\hspace{0.25cm}
\begin{math}
{\bf J}=\left(\begin{array}{cc}
1&0\\
0&-1
\end{array}\right)
\end{math},
\hspace{0.25cm}
\begin{math}
\\
{\bf \Lambda}=\left(\begin{array}{cc}
\lambda_1&0\\
0&\lambda_2
\end{array}\right)
\end{math},
\hspace{0.25cm}
\begin{math}
{\bf P}=\left(\begin{array}{cc}
0&\sqrt{a}Q(x,t)\\
-\sqrt{a}Q^*(x,t)&0
\end{array}\right)
\end{math},
\\\\
\begin{math}
{\bf W}=\left(\begin{array}{cc}ia|Q(x,t)|^2&\sqrt{a}\lambda x g(t)Q(x,t)+i\sqrt{a}Q_x(x,t)\\
-\sqrt{a}\lambda x g(t)Q^*(x,t)+i\sqrt{a}Q^*_x(x,t)&-ia|Q(x,t)|^2
\end{array}\right)
\end{math},\\
and $\zeta(t)=exp{\left(\int{\lambda g(t)dt}\right)}$.\\The constants $\lambda_1$ and $\lambda_2$ are arbitrary
constants. The subscripts $x$ and $t$ denote partial derivatives with respect to $x$ and $t$, respectively. The
function $Q(x,t)$ is related to the wave function through $Q(x,t)=\psi(x,t)\sqrt{\zeta}e^{c_1+i\lambda gx^2/4}$.
Equation~(\ref{gp2}) is obtained from the consistency condition $\Psi_{xt}=\Psi_{tx}$.

The linear system of 8 equations, Eqs.~(\ref{psi_x}) and (\ref{psi_t}), reduces to an equivalent system of 4
equations with nontrivial solutions by making the following substitutions: $\lambda_1=-\lambda_2^*$,
$\psi_2=\phi_1^*$, and $\phi_2=-\psi_1^*$. To be able to solve this reduced linear system we need to know an
exact ({\it seed}) solution of Eq.~(\ref{gp2}). Following our previous approach of finding seed
solutions~\cite{usama_darboux}, we find the following seed solution to Eq.~(\ref{gp2}) \begin{equation}
\psi_0(x,t)=\exp{\left[c_2+\lambda g/2+i\left(e^{\lambda g}k_0x-{2A^2a-k_0^2\over2\lambda}-{1\over4}\lambda{\dot
g}x^2+2A^2a\int{\left(e^{2(c_1+c_2+\lambda g)}+k_0^2e^{2\lambda g}\right)dt}\right)\right]}~, \label{seed}
\end{equation}
where $c_2$, $A$ and $k_0$ are arbitrary constants and ${\dot g}=dg/dt$.

The Darboux transformation can now be applied to the linear system
to generate a new solution of Eq.~(\ref{gp2}) as follows
\cite{salle}
\begin{equation}
\psi(x,t)=\psi_0(x,t)+2(\lambda_1+\lambda_1^*)e^{-c_1-i\lambda
gx^2/4+(\lambda/2)\int{gdt}}{\phi_1\psi_1^*\over(|\phi_1|^2+|\psi_1|^2)} \label{psinew}.
\end{equation}
Solving the linear system (\ref{psi_x}) and (\ref{psi_t}) using the seed solution Eq.~(\ref{seed}) and then
substituting for $\psi_0(x,t)$, $\psi_1(x,t)$, and $\phi_1(x,t)$ in the last equation, we obtain the following
new exact solution to Eq.~(\ref{gp2}):
\begin{widetext}
\begin{eqnarray}
\psi(x,t)&=&{\dot\eta}^{1/4}e^{-i{\ddot\eta}x^2/8{\dot\eta}} \left\{\frac{}{}
Ae^{q_1}\right.\nonumber\\&+&4\lambda_{1r}e^{i\theta_5+q_2}(2ic_3A\sqrt{a}e^{2\alpha\eta}+c_4q_3e^{-2i\theta_1+\Delta_r\sqrt{\dot\eta}x})(c_3q_3^*e^{2\alpha\eta}-2iA\sqrt{a}c_4e^{-i\theta_4+\Delta_r\sqrt{\dot\eta}x})\nonumber\\
&/&\left[c_3^2\Gamma e^{2\alpha\eta-\Delta_r\sqrt{\dot\eta}x}+c_4^2\Gamma
e^{-2\alpha\eta+\Delta_r\sqrt{\dot\eta}x}-4Ac_1c_2\sqrt{a}[(2\lambda_{1r}-\Delta_r)(\cos{2\theta_1}
+\cos{\theta_4})\right.\nonumber\\&+&\left.(\Delta_i-2\lambda_{1r})(\sin{2\theta_1}+\sin{\theta_4})+(\sin{2\theta_1}+\sin{\theta_4})k_0)
\frac{}{}\right]\left.\frac{}{}\right\} \label{gpsol},
\end{eqnarray}
\end{widetext}
where\\
\begin{math}
q_1=i[2A^2a(2\lambda\eta-1)+k_0(k_0(1-2\lambda\eta)+2\lambda\sqrt{\dot\eta})x]/2\lambda
\end{math},\\
\begin{math}
q_2=-2\alpha\eta-\Delta\sqrt{\dot\eta}x
\end{math},\\
\begin{math}
q_3=\Delta_i+k_0+i(\Delta_r+2i\lambda_{1i}-2\lambda_{1r})
\end{math},\\
\begin{math}
\theta_1=\Delta_r\lambda_{1r}\eta-((2\lambda_{1i}+k_0)-\sqrt{\dot\eta})\Delta_i/2
\end{math},\\
\begin{math}
\theta_2=(k_0^2(1-2\lambda\eta)-4\lambda\Delta_r\lambda_{1r}\eta)/4\lambda+2A^2a(1-2\lambda\eta)+2\Delta_i\lambda(2\lambda_{1i}\eta-\sqrt{\dot\eta}x)+2\lambda
k_0(\Delta_{i}\eta+\sqrt{\dot\eta}x)
\end{math},\\
\begin{math}
\theta_3=(2A^2a(2\lambda\eta-1)+k_0(k_0-2\lambda k_0\eta+2\lambda\sqrt{\dot\eta}x))/2\lambda
\end{math},\\
\begin{math}
\theta_4=(2\Delta_i\lambda_{1i}-2\Delta_r\lambda_{1r}+\Delta_ik_0)\eta-\sqrt{\dot\eta}\Delta_ix
\end{math},\\
\begin{math}
\theta_5=k_0\sqrt{\dot\eta}x+(2A^2a-k_0^2)(2\lambda\eta-1)/2\lambda
\end{math},\\
\begin{math}
\Gamma=(\Delta_i-2\lambda_{1i})^2+(\Delta_r-2\lambda_{1r})^2+4A^2a +k_0(2\Delta_i-4\lambda_{1i}+k_0)
\end{math},\\
\begin{math}
\alpha=\Delta_r\lambda_{1i}+\Delta_i\lambda_{1r}+\Delta_rk_0/2
\end{math},\\
\begin{math}
\Delta_r={\rm Re}[\sqrt{(2\lambda_1-ik_0)^2-4A^2a}]
\end{math},\\
\begin{math}
\Delta_i={\rm Im}[\sqrt{(2\lambda_1-ik_0)^2-4A^2a}]
\end{math},\\
\begin{math}
\gamma=\Delta_i+k_0+i(\Delta_r+2i\lambda_{1i}-2\lambda_{1r})
\end{math},\\
\begin{math}
\eta=\int{e^{2\lambda g}dt}
\end{math},\\
the subscripts $r$ and $i$ denote real and imaginary parts, respectively, and $c_{3}$ and $c_4$ are arbitrary
constants. Further reduction and simplification of this new general solution as performed in \cite{usamapre}
will not be considered here. Furthermore, we do not attempt to obtain all classes of solitonic solutions.
Instead, the focus in this paper will be on the effect of the parameters of the oscillating trapping potential
on the dynamics of single and multiple solitons which are described by Eq.~(\ref{gpsol}).

\section{\it Special case: Solitons in an oscillating harmonic trap}
\label{results}
\begin{figure}
\begin{center}
\includegraphics[width=10cm]{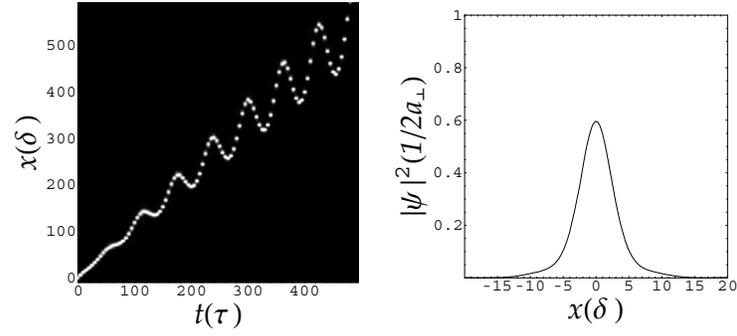}
\end{center}
\caption{(Left) Spatiotemporal contour plot of solitons density profile. (Right) Soliton density profile at
$t=0$. The values of the parameters used in this plot are: $-c_2=c_1=10$, $c_3=c_4=1$, $\lambda_{1i}=2$,
$\lambda_{1r}=1$, $A=0.1$, $a=0.2$, $k_0=0$, $\lambda=1$, $\omega=0.1$, $\delta=0$, $\alpha_1=-4$,
$\alpha_2=0.3$.} \label{fig1}
\end{figure}

\begin{figure}
\begin{center}
\includegraphics[width=5cm]{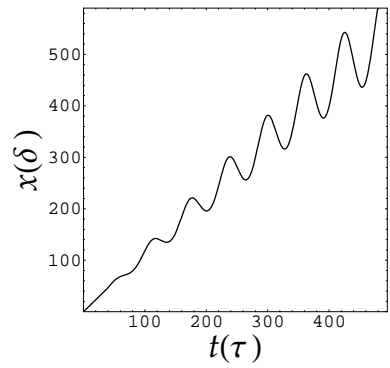}
\end{center}
\caption{Trajectory of the soliton peak density. The values of the parameters used here are the same as those of
Fig.1.} \label{fig2}
\end{figure}

In this section we derive from the general solution found in the previous section solitonic
solutions for an oscillating harmonic trapping potential. To this end, we take
\begin{equation}
\eta(t)=\int{e^{\alpha_1+\alpha_2\sin{(\omega t+\delta)}}dt}~,
\end{equation}
where $\alpha_1$, $\alpha_2$, $\omega$ , and $\delta$ are constants. Using the last equation and
$\eta=\int{e^{2\lambda g}dt}$ to substitute for $g$ in Eq.~(\ref{gp2}), the Gross-Pitaevskii equation takes the
form
\begin{equation}
i{\partial\psi(x,t)\over\partial t}=\left\{-{\partial^2\over\partial x^2}-{1\over8}\alpha_2\omega^2
\left[\sin{(\omega t+\delta)}+{1\over2}\alpha_2\cos{(\omega
t+\delta)^2}\right]x^2-2ae^{2c_1+(\alpha_1+\alpha_2\sin{\omega t+\delta})/2}|\psi(x,t)|^2\right\}\psi(x,t)
\label{gp3}.
\end{equation}
It should be noted that the specific choice of $\eta(t)=\exp{(2\lambda t)}$ leads to the Gross-Pitaevskii
equation of Ref.~\cite{liang} with an expulsive harmonic potential. Simulating the time dependencies of the
harmonic potential and the interatomic interaction of Eq.~(\ref{gp3}) may be difficult experimentally, but if we
choose the parameters such that $c_1\gg\alpha_1\gg\alpha_2$, the interatomic interaction can be considered
practically as constant, and the strength of the harmonic potential will be oscillating as $\sin{(\omega
t+\delta)}$. It should be noted, however, that even in this limit, the contribution of the positive $\cos(\omega
t+\delta)^2$ term will result in that the trapping potential will spend more time being an inverted parabola
than being an upright parabola. Over a large enough time interval, this will lead to expelling the solitons away
from the center of the trap. This is indeed the picture that we get when we plot the density of a single-soliton
solution as shown in Fig.~\ref{fig1}. It is clear from this figure that the soliton is being expelled out from
the center of the condensate at $x=0$. The oscillation in the trajectory of the soliton is due to the
oscillating trapping potential. The discontinuous appearing of the soliton's peak density is due to the
interaction with the background. This trajectory can be also extracted from the general solution
Eq.~(\ref{gpsol}) by considering the term $(\Gamma c_3^2e^{2\alpha\eta-\sqrt{\dot\eta}\Delta_rx}+\Gamma
c_4^2e^{-2\alpha\eta+\sqrt{\dot\eta}\Delta_rx})$ in the denominator. At the peaks of the oscillation in
Fig.~(\ref{fig1}), this is the dominant term that determines the position of the peak of the soliton.
Specifically, the peaks are given by the condition $2\alpha\eta-\sqrt{\dot\eta}\Delta_rx=0$. Using this
condition to plot $x$ vs. $t$ in Fig.~\ref{fig2}, we obtain a curve that is identical to the soliton trajectory
in Fig.~\ref{fig1}. The mean slope of this curve is proportional to $\alpha/2\Delta_r$. Hence, the rate at which
the soliton leaves the center of the trapping potential can be delayed by choosing the parameters and the
arbitrary constants such that $\alpha/2\Delta_r$ is small.

For the special case of $\alpha=0$ the center of mass of the soliton will be located at $x=0$ at all times. The
oscillating trapping potential will result only in oscillations in the width and peak density of the solitons.
The solitons in this case remain trapped at the center of the trapping potential. We show this case in
Figs.~\ref{fig3}-\ref{fig6} where we see a multi-soliton solution with central soliton being {\it pinned} at
$x=0$ and off-central ones oscillating around their initial positions. In Fig.~\ref{fig4}, we use the same
parameters as in Fig.~\ref{fig3} but with a doubled frequency $\omega$. This shows clearly that the oscillations
of the off-centered solitons are due to the oscillation in the trapping potential. In Fig.~\ref{fig5}, we show
that for some values of the parameters the dynamics of the peak soliton density can be drastic such that the
soliton disappears in the background and reappears at regular discrete times. Figure~\ref{fig6} is similar to
Fig.~\ref{fig3} but with a larger number of solitons.

\section{conclusions}
\label{conc} 
\begin{figure}
\begin{center}
\includegraphics[width=10cm]{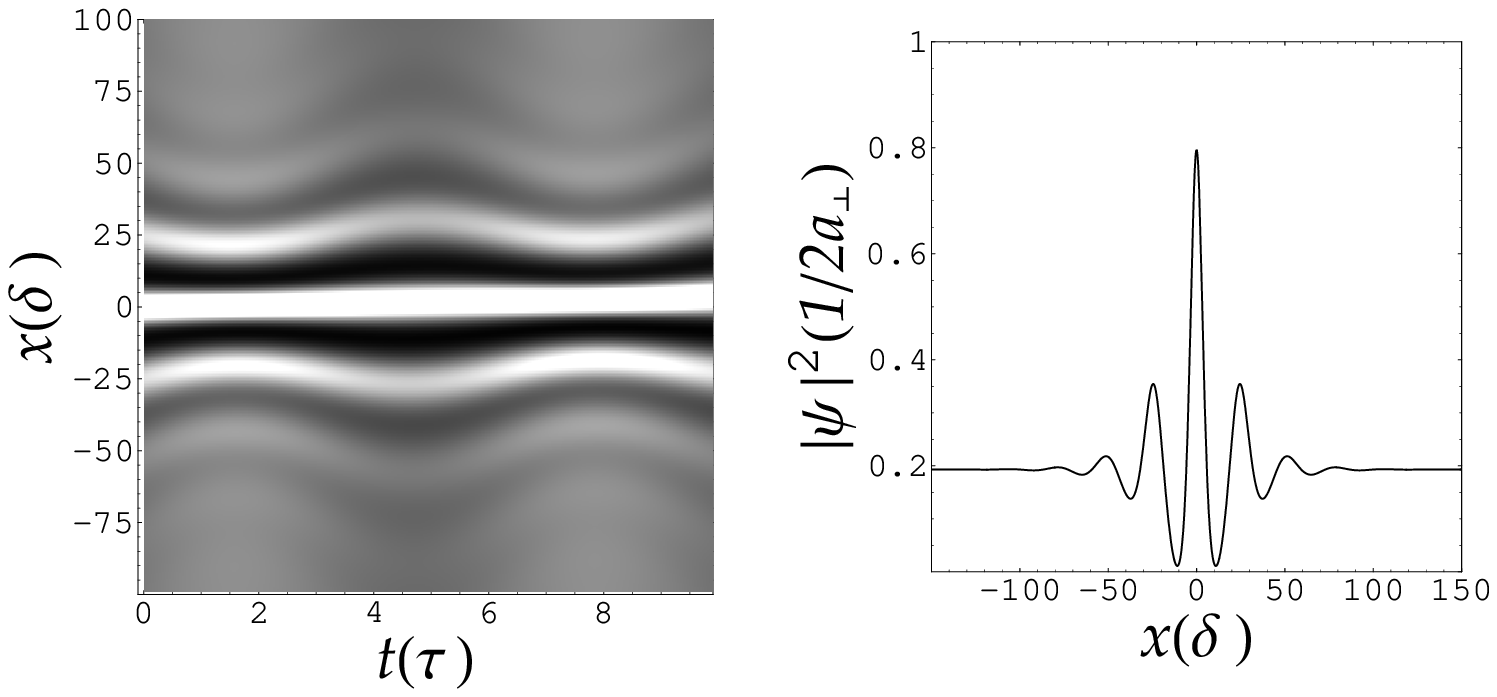}
\end{center}
\caption{(Left) Spatiotemporal contour plot of solitons density profile. (Right) Soliton density profile at
$t=0$. The values of the parameters used in this plot are: $-c_2=c_1=10$, $c_3=c_4=1$,
$\lambda_{1i}=\lambda_{1r}=1$, $A=2$, $a=0.9$, $k_0\sim5.03$, $\lambda=\omega=1$, $\delta=0$, $\alpha_1=-6$,
$\alpha_2=0.3$. The value of $k_0$ is the solution of $\alpha=0$ with respect to $k_0$.} \label{fig3}
\end{figure}

\begin{figure}
\begin{center}
\includegraphics[width=10cm]{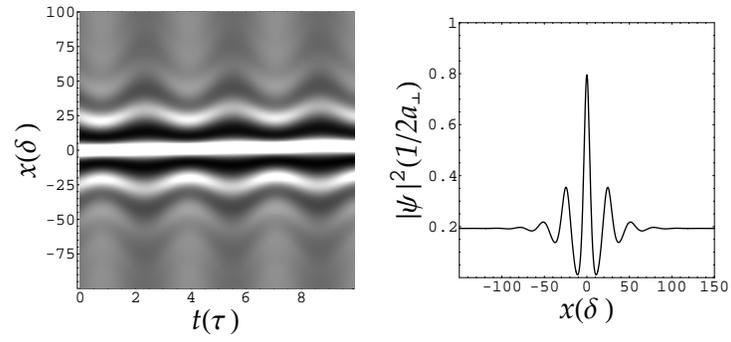}
\end{center}
\caption{Spatiotemporal contour plot of density solitons density profile. The values of the parameters used in
this plot are the same as those of Fig.3 but with $\omega=2$.} \label{fig4}
\end{figure}

\begin{figure}
\begin{center}
\includegraphics[width=10cm]{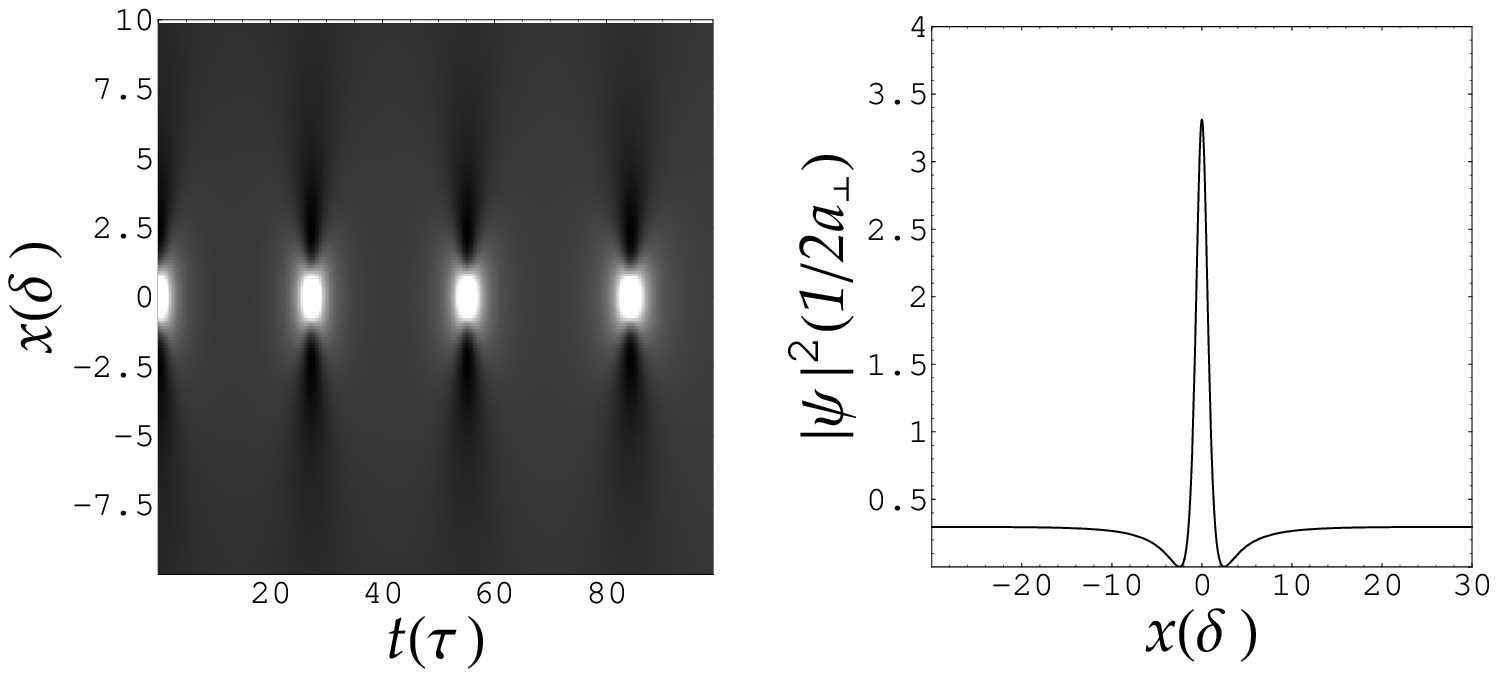}
\end{center}
\caption{(Left) Spatiotemporal contour plot of solitons density profile. (Right) Soliton density profile at
$t=0$. The values of the parameters used in this plot are: $-c_2=c_1=10$, $c_3=c_4=1$, $\lambda_{1i}=0$,
$\lambda_{1r}=1$, $A=2$, $a=0.9$, $k_0=0$, $\lambda=1$, $\omega=0.01$, $\delta=0$, $\alpha_1=-2$,
$\alpha_2=0.3$.} \label{fig5}
\end{figure}

\begin{figure}
\begin{center}
\includegraphics[width=10cm]{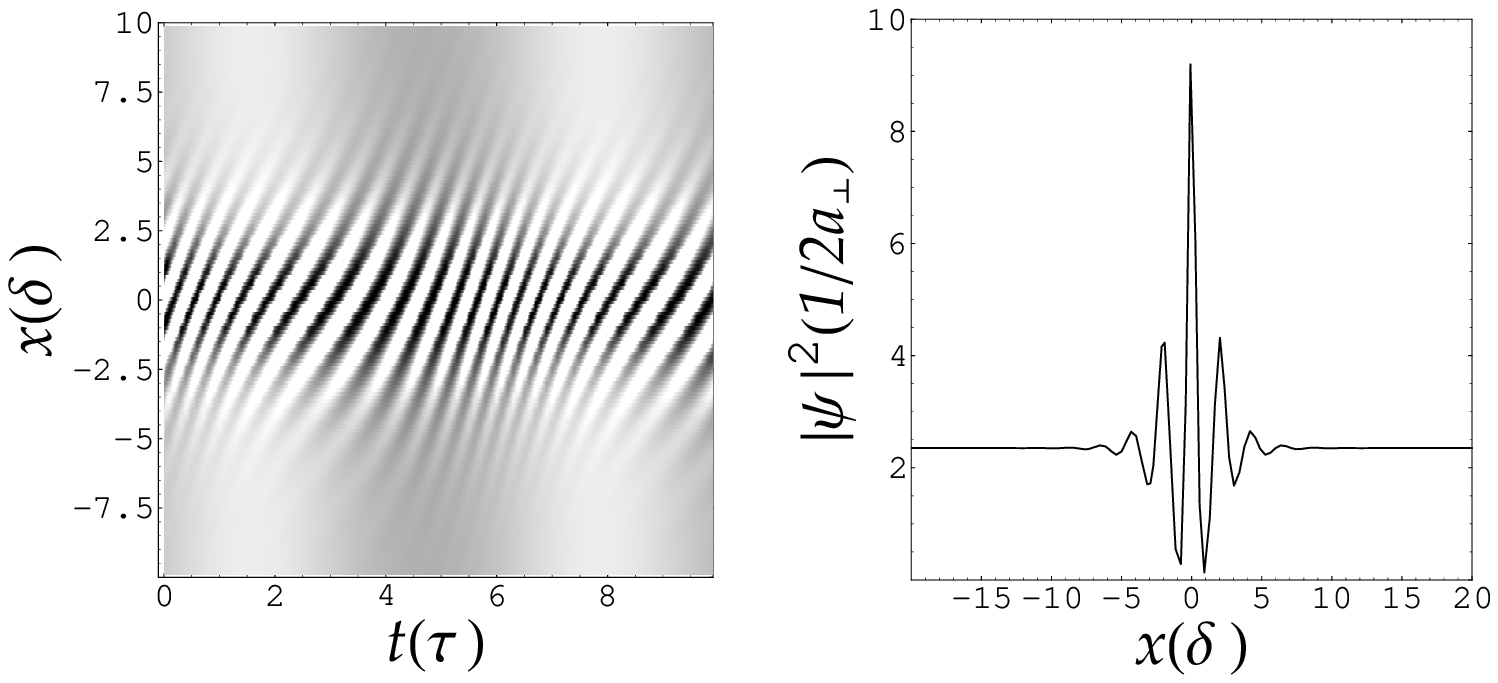}
\end{center}
\caption{(Left) Spatiotemporal contour plot of solitons density profile. (Right) Soliton density profile at
$t=0$. The values of the parameters used in this plot are the same as those of Fig.3 but with $\alpha_1=-1$.}
\label{fig6}
\end{figure}

We have found exact single and multi solitonic solutions of a time-dependent Gross-Pitaevskii
equation with time-dependent amplitudes of harmonic trapping potential and interatomic interaction. We
considered the interesting special case of sinusoidally oscillating strength of the harmonic trapping potential.
We focused on the effect of the frequency and amplitude of the oscillating trapping potential on the dynamics of
the solitons. The parameters can be chosen such that the soliton is trapped at the center of the condensate. For
a typical $^{87}$Rb condensate with $10^5$ atoms and trapping frequency of order 100 Hz, the Thomas-Fermi size
of the condensate is $R_{TF}= (Na/\delta)^{1/5}\delta\approx 10\delta$ \cite{pethickbaym}. This means that the
width of the solitons of Figs.~\ref{fig3}-\ref{fig4} are of the order of the size of a nonsolitonic condensate
while the width of the solitons in Figs~\ref{fig5}-\ref{fig6} is smaller than the size of the nonsolitonic
condensate.


\begin{thebibliography}{99}


\bibitem{burger} S. Burger, K. Bongs, S. Dettmer,
W. Ertmer, and K. Sengstock, Phys. Rev. Lett. {\bf 83}, 5198 (1999).

\bibitem{denschlag} J. Denschlag {\it et al.},
Science {\bf 287}, 97 (2000).

\bibitem{anderson} B.P. Anderson, P.C. Haljan,
C.A. Regal, D.L. Feder, L.A. Collins,
C.W. Clark, and E.A. Cornell, 
Phys. Rev. Lett. {\bf 86}, 2926 (2001).

\bibitem{schreck} L. Khaykovich {\it et al.},
Science {\bf 296}, 1290 (2002).

\bibitem{randy} K.E. Strecker, G.B. Partridge, A.G. Truscott,
and R.G. Hulet, Nature {\bf 417}, 150 (2002).

\bibitem{eirman} B. Eiermann {\it et al.},
Phys. Rev. Lett. {\bf 92}, 230401 (2004).


\bibitem{linear1}
U. Al Khawaja, {\it et al.}, Phys. Rev. Lett., {\bf 89}, 200404
(2004).

\bibitem{linear2}
L. D. Carr and J. Brand, Phys. Rev. Lett., {\bf 92}, 040401 (2004).



\bibitem{cast} L.D. Carr and Y. Castin, Phys. Rev. {\bf 66}, 063602 (2002).


\bibitem{busch} Th. Busch and J.R. Anglin, Phys. Rev. Lett. {\bf 84}, 2298 (2000); L.
Salasnich, Phys. Rev. A {\bf 70}, 053617 (2004).


\bibitem{abdu} F. K. Abdullaev, A. Gammal, A. Kamchatnov, L. Tomio,
Int. Jour. of Mod. Phys. B, {\bf 19}, 3415 (2005).



\bibitem{sala}L. Salasnich, A. Parola, and L. Reatto, Phys. Rev. Lett. {\bf91},
080405 (2003); K. Kasamatsu and M. Tsubota, ibid. {\bf93}, 100402 (2004).

\bibitem{usamapre} U. Al Khawaja, to appear in Phys. Rev. E.



\bibitem{wu} Wu-Ming Liu, B. Wu, and Q. Niu, Phys. Rev. Lett. {\bf
84}, 2294 (2000).

\bibitem{jun}  J. Ieda, T. Miyakawa, and M. Wadati
Phys. Rev. Lett. {\bf 93}, 194102 (2004).

\bibitem{liang} Z.X. Liang, Z.D. Zhang, and W.M. Liu,
Phys. Rev. Lett. {\bf 94}, 050402 (2005).

\bibitem{lu} Lu Li, Zaidong Li, B.A. Malomed, D. Mihalache, and W.M. Liu
Phys. Rev. A {\bf 72}, 033611 (2005).

\bibitem{raj} R. Atre, P.K. Panigrahi, and G.S. Agarwal,Phys. Rev. E {\bf 73},
056611(2006).



\bibitem{usama_darboux} U. Al Khawaja, J. Phys. A: Math. Gen. {\bf 39}, 9679 (2006).


\bibitem{salle} V.B. Matveev and M.A. Salle, {\it Dardoux Transformations and
Solitons}, Springer Series in nonlinear Dynamics (Springer-Verlag,
Berlin, 1991).




\bibitem{fesh} S. Inouye, M.R. Andrews, J. Stenger, H.-J. Miesner,
D.M. Stamper-Kurn, and W. Ketterle, Nature (London) {\bf 392}, 151 (1998); S.L. Cornish, N.R. Claussen, J.L.
Roberts, E.A. Cornell, and C.E. Weiman, Phys. Rev. Lett. {85}, 1795 (2000).



\bibitem{carr} L.D. Carr, C.W. Clark, and W.P. Reinhardt,
               Phys. Rev. A. {\bf 62}, 063610 (2000);
               L.D. Carr, C.W. Clark, and W.P. Reinhardt,
               Phys. Rev. A. {\bf 62}, 063611 (2000).


\bibitem{note} This is performed by inspection.
Equivalently, one can use the method developed in \cite{usama_darboux}.

\bibitem{pethickbaym} G. Baym and C.J. Pethick, Phys. Rev. Lett. {\bf 76}, 6 (1996).



\end{thebibliography}
\end{document}